\documentstyle[11pt]{article}

\begin{document}

\setcounter{page}{0}

\title{
\rightline{\small{hep-ph/0205100}} \vspace*{0.5cm}
\bf {B decays and Supersymmetry} \footnote{Working group talk presented
at WHEPP-7, HRI, Allahabad, India, Jan. 2002. To be published in Pramana} }
\author{ {\large\bf Anirban Kundu}\thanks{Electronic address:
akundu@juphys.ernet.in}\\[3mm]
Department of Physics, Jadavpur University, Kolkata 700032, India}

\date{\today}
\maketitle



\begin{abstract}
I discuss how supersymmetry affects various observables
in B decays, and point out the interesting
channels in the context of B factories.
\end{abstract}

\section{Introduction}
Supersymmetry (SUSY) does not need any motivation any more. 
It is, however, interesting to ask whether one can have indirect
SUSY signals from low-energy observables before LHC starts. 
The answer is {\em yes}, and in
this talk we point out some of the SUSY signals that one may observe at
the leptonic and hadronic B-factories. 
These factories have already started taking data, and in the near 
future, some of
the observables from the factories will attain a precision which should be
able not only to test the Standard Model (SM) but also to probe for
physics beyond the Standard Model (BSM). In fact with a little bit of luck
the B factories may be the first place where one will see the BSM signals
much before the LHC starts. Such signals could then be enhanced and studied
in detail in hadronic B-factories or in very high luminosity $e^+e^-$
experiments like SuperBaBar \cite{hitlin}. Before going into such signals, 
let us first see why we need this `luck'.

The present generation $e^+e^-$ B factories will in no way verify the CKM
ansatz of CP violation. They will measure quantities like $V_{cb}$, $V_{ub}$,
$(\Delta M/\Gamma)_{B_d}$, $\sin 2\beta$ and $\sin 2\alpha$ to various degrees
of precision. What one can at most say is that all measurements are 
consistent with the CKM picture, but existence of BSM cannot be ruled out,
except that its parameter space will become more and more constrained with
increasing amount of data. In other words, these indirect measurements
will not be able to see the signatures of any arbitrary SUSY model; 
the model must
have the parameters lying in the right ballpark. This statement will be
quantified later.

All SUSY models available in the market can be divided into two broad
categories: R-parity conserving (RPC) and R-parity violating (RPV), where
$R=(-1)^{3B+L+2S}$. The RPC models are further divided according to how
SUSY breaks: the supergravity (SUGRA) type, the gauge-mediated (GMSB) type,
the anomaly-mediated (AMSB) type, and so on \cite{probir}. In these models
the SUSY effects always appear in loops and are therefore harder to detect
unless the corresponding SM process is either absent or loop-mediated itself.
Thus, one does not expect to see a SUSY signal in tree-level $b\rightarrow c,u$
decays, but $B$-$\bar{B}$ mixing, $b\rightarrow s\gamma$ or $b\rightarrow
K(^*)\ell^+\ell^-$  are good places to look for such signals.

RPV SUSY is different in the sense that one can have tree-level slepton or
squark mediated B decays and thus the new physics amplitude can very well
compete with, or even overwhelm, the SM amplitude. The expected CP asymmetry
in a particular decay may be completely different from the SM expectation;
the branching ratio (BR) can be substantially high or low from the SM
value; even the SM forbidden modes may appear. For lack of space we will just
discuss some of the most important effects coming from RPV SUSY in B decays.

\section{FCNC in SUSY}
In R-parity conserving SUSY models, there can be two more independent
phases $\phi_A$ and $\phi_B$ apart from the CKM phase, but the electric dipole
moment of neutron constrains both of them to be $\sim{\cal O}(10^{-2}$-
$10^{-3})$ unless the squarks are extremely 
heavy or there is a fine-tuning between
$\phi_A$ and $\phi_B$ \cite{ellis}. 
We take both of them to be zero, a choice which can be 
theoretically motivated, since $\phi_A$ ($\phi_B$) is the relative
phase between $A$ ($B$) and the common gaugino mass $M$ \cite{phase}. 
Even then one can have new contribution to 
CP-violation coming from SUSY FCNC effects \cite{susyfcnc}. 
The origin of SUSY FCNC can be
easily understood: quark and squark mass matrices are not simultaneously
diagonalizable. 
%
At $q^2\sim m_W^2$, radiative corrections induced by up-type (s)quark loops
are important. These corrections are typically of the order of $\log(\Lambda_S/
m_W)$ and hence can be large for SUGRA type models. This generates FCNC which
occurs even in the quark-squark-neutral gaugino vertices, but the flavour
structure is controlled by the CKM matrix (this need not be true in any
arbitrary SUSY model). 

One generally works in the basis where the quark fields are eigenstates of
the hamiltonian. SUSY FCNC can be incorporated in two ways: (i) {\em vertex 
mixing}, an approach where the squark propagators are flavour and `chirality'
conserving, and the vertices violate them; (ii) {\em propagator mixing},
where flavour and `chirality' are conserved in vertices but changed in
propagators. The second approach is more preferred for phenomenological
analysis, since the higher order QCD corrections are known there. 

Thus, at the weak scale one can write the $6\times 6$ squark mass matrix
(say the down-type) as
\begin{equation}
\tilde{{\cal M}_D}^2=\pmatrix{{\cal M}_{DLL}^2|^{tree}+\Delta_{LL}^2 &
\Delta_{LR}^2\cr \Delta_{RL}^2 & {\cal M}_{DRR}^2|^{tree}+\Delta_{RR}^2}
\end{equation}
where the $\Delta$ terms incorporate the FCNC effects. In fact, only 
$\Delta_{LL}^{ij}$ changes flavour, but the other $\Delta^{ij}$s are not
\cite{gabbiani}. Different FCNC effects are parametrized in terms of 
$\delta\equiv\Delta/\tilde m^2$, where $\tilde m=\sqrt{\tilde m_1\tilde m_2}$.
Of course they are completely calculable in MSSM; all such $\delta$s are
$\leq {\cal O}(10^{-2})$. This is not true in general SUSY models; and
such models can be constrained by the values of $\delta$ they produce. 
Theoretically, for the success of perturbative analysis, one expects
$|\delta|<1$.

The standard way to look for SUSY models whose signatures can be found
in low-energy machines can thus be divided in two steps:
(i) Compute the constraints on $Re$ and $Im$ $\delta$ from low-energy
observables like $K-\bar K$, $B-\bar B$ and $B_s-\bar{B_s}$ mixing, 
$\epsilon_K$, $\epsilon'/\epsilon$, $b\rightarrow s\gamma$ etc. Unconstrained
$\delta$s can have $|\delta|\leq 1$ and arbitrary phases.
(ii)Find CP-conserving and CP-violating effects compatible with such
constraints.

Before proceeding further, let us note a few points.
Since there is no a priori reason why the FCNC should be small, there
must be some inherent mechanism, from a theoretical point, that suppresses 
FCNC. This can be {\em alignment}, where quark and squark mass matrices
are aligned even at the electroweak scale; {\em heavy squarks}, where the
squarks --- particularly those in the first two generations --- are at
the TeV scale; or some {\em family symmetry} which suppresses such FCNC.
Since SUSY particles always appear in loops, the leading contribution
is a constant (the so-called superGIM mechanism), and effects are observable
only if the corresponding SM amplitude is zero or loop-induced. Furthermore,
SUSY does not induce any new operators from that in the SM, so all one has 
to do is to compute the SUSY Wilson coefficients. 

Recently, it has been shown \cite{demir} that even if the SUSY phases are
large, there cannot be any significant CP-violating effect if the flavour
structure is governed by the CKM matrix alone (the universal unitarity 
triangle scenario). This means that B-factories will probe the flavour
structure of SUSY much before LHC! However, this is not true for general
SUSY flavour models, and we will concentrate on those models where this
constraint can be bypassed.

\section{$B_d$-$\bar{B_d}$ mixing, Unitarity Triangle and SUSY}

In the SM, the phase coming from $B_d$-$\bar{B_d}$ is $2\beta$. It
can be shown that all four new boxes ($t$-$H^+$, $\tilde t$-$\tilde \chi^+$,
$\tilde b$-$\tilde g$ and $\tilde b$-$\tilde \chi^0$) coming in MSSM
with alignment have same phases and hence the CP asymmetry does not
change. 

In general SUSY models the off-diagonal SUSY hamiltonian 
$M_{12}^{SUSY}$ depends on $(\delta^d_{13})_{LL,LR,RR}$ and their phases
may change the prediction of $\sin 2\beta$ by 10-15\% \cite{bbbar}. 
Thus, the measured
values of $\beta$ and $\alpha$ change, but by a compensating amount if the
former is extracted from $B_d\rightarrow J/\psi K_S$ (or $\phi K_S$) and
the latter from $B_d\rightarrow \pi^+\pi^-$ (since one actually measures
$\beta+\gamma=\pi-\alpha$):
\begin{equation}
\beta' = \beta+\phi_{SUSY},\ \ \alpha' = \alpha-\phi_{SUSY}
\end{equation} 
so that their sum remains unchanged. $\gamma$ measured from $B_s$ decays
will also be changed by a different amount, and hence the unitarity
triangle (UT) will not close: this is the signal. However, $\gamma$
determination needs hadronic machines and is not easy, and even $\alpha$
determination is substantially contaminated with penguins as is evident
from the recent CLEO, BaBar and BELLE data \cite{bpipi}.

\section{$b\rightarrow s\gamma$ : the most `promising' place to find SUSY?}

The radiative penguin decay $b\rightarrow s\gamma$, and $b\rightarrow
s\ell^+\ell^-$, which is closely related to it, has been discussed in the
literature in great detail \cite{btos}. 
The CP asymmetry in $b\rightarrow s\gamma$ is
negligible in the SM. The SUSY contribution, on the otherhand, is zero if
SUSY is unbroken \cite{bg}. This decay is controlled by the magnetic penguin
operator $O_7 \sim \bar s\sigma_{\mu\nu}(m_bP_R+m_sP_L)bF^{\mu\nu}$.

In MSSM only the electroweak $t$-$H^+$ and $\tilde t$-$\tilde \chi^+$
penguins are important (gluino and neutralino penguins are subdominant,
and vanish if squark mass matrices are flavour-diagonal). The $t$-$H^+$
penguin always adds constructively with the SM penguin; the stop-chargino
penguin can be either constructive or destructive depending on the mass
and composition of stop and chargino. Of course, the effect is significant
only for relatively light stop and chargino: the BR can even be doubled 
if they are $\sim 100$ GeV. For this particular parameter space, enhancement
of BR is a good signal, more so if one remembers that the theoretical
prediction for SM and the experimental numbers are in the same ballpark. 

For destructive interference, there can be another interesting
observable \cite{bartl}: the CP asymmetry $A_{CP}^{b\rightarrow s\gamma}$,
defined as
\begin{equation}
A_{CP}^{b\rightarrow s\gamma} = {BR(\bar B_d\rightarrow X_s\gamma) -
BR(B_d\rightarrow X_s\gamma)\over BR(\bar B_d\rightarrow X_s\gamma) + 
BR(B_d\rightarrow X_s\gamma)}.
\end{equation}
This can go upto 5\% for destructive $\tilde t$-$\tilde\chi^+$ loop
where the BR is about the lowest possible experimental value: $2\times
10^{-4}$. For constructive loops, this asymmetry quickly falls to zero.
This signal can just be observable in current $e^+e^-$ factories and
will certainly be observable in future high-lumonisty machines.

In general SUSY models, the data on $b\rightarrow s\gamma$ constrains
$|(\delta^d_{23})_{LR}| \leq 1.6\times 10^{-2}$ for $m_{\tilde g}=m_{\tilde
q}$. Corresponding $LL$ term is not at all constrained since the chirality
flip occurs on the $b$ quark line and not on the $\tilde g$ line.

\section{Leptonic and Semileptonic $B$ decays}

Recently Belle has observed $B\rightarrow K\mu^+\mu^-$; all three
$e^+e^-$ machines have put constraints on other $B\rightarrow K(^*)
\ell^+\ell^-$ modes \cite{belle-kmumu}. Unfortunately the SM prediction
for these exclusive modes is not very pecise \cite{alietal}. The
exclusive mode $B\rightarrow X_se^+e^-$ ($B\rightarrow X_s\mu^+\mu^-$)
can be changed by a factor of 40-500\% (25-550\%) in general SUSY
models. The forward-backward asymmetry $A_{FB}(e/\mu)$ can range between
$-0.18$ to $0.33$ in contrast to a SM prediction of $0.23$. These signals
should be measurable in current and upcoming factories.

The dilepton and single lepton asymmetries may also be useful tools
for distinguishing SUSY flavour models. However, this definitely requires
hadronic or high-luminosity $e^+e^-$ machines \cite{su-randall}. 

\section{Nonleptonic $B$ decays}

These processes are plagued mainly with hadronic uncertainties. As stated
already, if the flavour structure is governed by the CKM matrix, the CP
violation is bound to be small. In general SUSY models, decays proceeding
through $b\rightarrow s$ have a competing chance since the SM process is
penguin and the corresponding $\delta$, $(\delta^d_{23})_{LL}$ is
unconstrained both in its magnitude and phase. For example, the exclusive
mode $B_d\rightarrow \phi K_S$ may have a SUSY amplitude which can be 
70\% (20\%) for $m_{\tilde q}=250$ ($500$) GeV. Thus, $A_{CP}$ from this
mode may deviate significantly from $\sin 2\beta$. The MSSM effect may be
just perceptible in $B_d\rightarrow J/\psi K_S$. There is almost no hope
for detecting RPC SUSY signals from other nonleptonic decays.

\section{R-parity violating SUSY and $B$ decays}

This is, in some sense, a goldmine for phenomenologists. 
From the superpotential
\begin{equation}
W={1\over 2}\lambda_{ijk}L_iL_jE^c_k+
\lambda'_{ijk}L_iQ_jD^c_k+
{1\over 2}\lambda''_{ijk} U^c_iD^c_jD^c_k,
\end{equation}
one constructs
a four-fermi effective theory by integrating out the slepton or squark fields
to find that (i) with slepton exchange, $\lambda\lambda'$ type products
contribute to leptonic and semileptonic decays; (ii) $\lambda'\lambda'$ 
products contribute to nonleptonic decays; (iii) with squark exchange,
$\lambda''\lambda''$ products contribute to nonleptonic decays.

There are two different approaches in the literature. First, one takes
a particular product coupling --- allowed with the phenomenological
constraints --- and finds the implication for different decay modes.
People have shown that with suitable values of RPV couplings, (i)
$A_{CP}$ in $B_d\rightarrow J/\psi K_S$ and in $B_d\rightarrow \phi K_S$
may be significantly modified \cite{guetta}; (ii) BR for the
mode $B\rightarrow \eta' K$ can be enhanced to explain the discrepancy
between data and prediction assuming no charm content in $\eta'$ \cite{cdk};
(iii) modes forbidden in SM (like $\Delta B=1, \Delta S=2$) may be seen
even in present-day colliders \cite{huitu-singer}; (iv) prediction for
$b\rightarrow s\gamma$ can be modified \cite{gg-chang}; (v) there
can be couplings which contribute to both neutral $B$ mixing and tree-level
decays, and thus the signal can be more complex \cite{adggak2}.

The second approach constrains various products from data on mixing and
decay to rare channels. Significant bounds on various $\lambda \lambda'$
and $\lambda'\lambda'$ products have been obtained from $B_d$-$\bar{B_d}$
mixing \cite{gg-arc}, $B_{d,s}\rightarrow \ell^+_i\ell^-_j$ \cite{kim,akjps},
$B\rightarrow K\ell_i^+\ell_j^-$ \cite{akjps}. These bounds generally are
orders of magnitude better than the bounds obtained from other processes.

\section{Conclusion}

The best possible searching grounds for RPC SUSY signals
are $b\rightarrow s\gamma$ and, to a certain extent, $b\rightarrow
s\ell^+\ell^-$. One should look for a change in BR and CP and/or FB
asymmetries.

RPV models have a rich phenomenology, not all of which have been explored
so far. Both BR and $A_{CP}$ can change significantly. To be more precise,
$A_{CP}$ coming from modes which are expected to yield same result for SM
may give completely different results. SM forbidden processes like
$B\rightarrow e\mu$ may be observed. 
%

\section{Acknowledgement}

I thank the organisers of WHEPP-7, particularly my long-time collaborator
Biswarup Mukhopadhyaya, for arranging a stimulating workshop. Thanks are
also due to those people with whom I collaborated about some of the topics
discussed here: Gautam Bhattacharyya, Debajyoti Choudhury, Amitava Datta,
Bhaskar Dutta and Jyoti Prasad Saha. This work was supported by BRNS,
India (Project No.\ 2000/37/10/BRNS) and UGC, India (Project No.\ F.10-14/2001
(SR-I)).

\end{document}